\documentclass[fleqn,usenatbib]{mnras}
\usepackage{newtxtext,newtxmath}
\usepackage{graphicx}	
\usepackage{amsmath}	

\usepackage{amssymb}	
\usepackage{bm}
\usepackage{amsfonts}
\usepackage{latexsym}
\usepackage{amsmath}
\usepackage{epsfig}
\usepackage{amsbsy}
\usepackage{enumitem}
\usepackage{multirow}
\usepackage[T1]{fontenc}
\usepackage{orcidlink}

\DeclareRobustCommand{\VAN}[3]{#2}
\let\VANthebibliography\thebibliography
\def\thebibliography{\DeclareRobustCommand{\VAN}[3]{##3}\VANthebibliography}

\title[Neutrino emission from ULXPs]{Polar Mounds on Strangeon Stars: the Neutrino Emission from Ultraluminous X-ray Pulsars}

\author[H.-B. Li et al.]{
Hong-Bo Li,$^{1}$\thanks{E-mail: lihb2020@pku.edu.cn}\orcidlink{0000-0002-4850-8351}
Shi-Jie Gao,$^{2,3}$\orcidlink{0000-0002-0822-0337}
Xiang-Dong Li,$^{2,3}$\orcidlink{0000-0002-0584-8145}
and Ren-Xin Xu$^{1,4}$\thanks{E-mail: r.x.xu@pku.edu.cn}\orcidlink{0000-0002-9042-3044}
\\
$^{1}$Kavli Institute for Astronomy and Astrophysics, Peking University, Beijing 100871, China\\
$^{2}$School of Astronomy and Space Science, Nanjing University, Nanjing, 210023, China\\
$^{3}$Key Laboratory of Modern Astronomy and Astrophysics, Nanjing University, Ministry of Education, Nanjing, 210023, China \\
$^{4}$Department of Astronomy, School of Physics, Peking University, Beijing 100871, China
}

\date{Accepted XXX. Received YYY; in original form ZZZ}

\pubyear{\the\year{}}

\begin{document}
\label{firstpage}
\pagerange{\pageref{firstpage}--\pageref{lastpage}}
\maketitle

\begin{abstract}
Ultraluminous X-ray pulsars (ULXPs) serve as unique astrophysical laboratories, offering critical insights into accretion physics under extreme conditions, such as strong magnetic fields and super-Eddington accretion rates. Additionally, the nature of pulsars, i.e., the equation of state of supranuclear matter, is still a matter of intense debate, basing on either conventional neutron stars or strange stars. In this work, in order to differentiate the conjectured states of matter, we investigate accretion columns of ULXPs based on the strangeon-star (SS) model, focusing on the thermal mound at the column base.  Accounting for Coulomb and strangeness barriers of SSs, we find that the mound can reach $0.7-0.95\,\rm km$ in height with temperatures above $10^9\, \rm K$, enabling substantial neutrino emission via electron-positron annihilation. At low accretion rates  ($< 10^{20}\, \rm g\,s^{-1}$), photons dominate the luminosity, while at higher rates ($> 10^{21}\, \rm g\, s^{-1}$), photon trapping makes neutrino emission the main cooling channel, with total luminosity exceeding photon emission, which saturates near $10^{41}\, \rm erg\,s^{-1}$. Even though the predicted neutrino flux from the nearest system, Swift J0243.6$+$6124, lies well below the diffuse MeV background--implying that detectable emission would require substantially closer or more luminous sources--these results demonstrate the key role of the thermal mound and SS properties in accretion, providing a foundation for future ULXP studies and suggesting that neutrino observations could, in principle, offer a novel probe of SSs and extreme supranuclear matter.
\end{abstract}


\section{Introduction}
\label{sec: intro}
 
 Ultraluminous X-ray sources (ULXs) are extragalactic, point-like X-ray emitters not associated with active galactic nuclei, exhibiting X-ray luminosities significantly exceeding $L_{\rm X} >  10^{39}\,\rm erg \, s^{-1}$ \citep{Kaaret:2017tcn, 2021AstBu..76....6F, King:2023nft}. Traditionally, ULXs were interpreted as systems powered either by super-Eddington accretion onto stellar-mass black holes \citep{Begelman:2006bi, Poutanen:2006uc} or by accretion onto intermediate-mass black holes \citep{Colbert:1999es, 2006ASPC..352..121M, Maccarone:2007dd, Koliopanos:2017aja}. However, the discovery of coherent X-ray pulsations in M82 X-2 \citep{Bachetti:2014qsa}, which exhibits an extraordinary luminosity of ${L_{\mathrm{X}} \simeq 1.4 \times 10^{40}\, \rm erg \, s^{-1}}$ and a pulse period of $1.37 \, \rm s$, provided compelling evidence that some ULXs are powered by accreting pulsars rather than black holes. In these systems, matter channeled by the strong magnetic field onto the stellar surface forms hot spots that produce pulsed X-ray emission as the star rotates.

Since the discovery of M82 X-2, numerous other ULX pulsars (ULXPs) have been identified, and their observed properties are summarized in Table \ref{tab: Table_1}. Of particular interest is the possible detection of a narrow cyclotron scattering feature at $\sim 4.5\, \rm keV$ in M51 ULX-8 \citep{Brightman:2018fsb}. Although pulsations have not been detected from this source, the feature has been interpreted as a proton cyclotron line, implying a surface multipole magnetic field as strong as $B \sim 10^{15} \, \rm G$. The X-ray luminosities of  ULXPs are typically in the  range of $10^{39}- 10^{41}\, \rm erg \, s^{-1} $, which is above the Eddington luminosity for accreting pulsars, $L_{\rm Edd} \approx 1.8 \times 10^{38}\,(M/M_{\odot}) \, \rm erg \, s^{-1}$, where $M$ is the mass of the compact object. This challenges our understanding of accretion physics under extreme magnetic fields, high temperatures, and super-Eddington regimes, making ULXPs unique laboratories for exploring high-energy astrophysical processes.

The origin of the extremely high luminosity observed in ULXPs remains a subject of ongoing debate. Proposed models for ULXPs span a broad range of surface magnetic field strengths. At relatively low fields ($\sim 10^{11}\, \rm G$), the observed luminosities may be explained by strong radiatively driven outflows and geometrical beaming \citep{Middleton:2016qss, 2017MNRAS.471L..71M, King:2019jvg, Lasota:2023xef}. At the other extreme, magnetar-like fields ($\sim 10^{14}\, \rm G $) can suppress radiation pressure \citep{1976MNRAS.175..395B, Tong:2014esa, DallOsso:2015nzq, Eksi:2014lya, Chandra:2025omf} and confine accretion columns near the magnetic poles \citep{Mushtukov:2015zea, 2020PASJ...72...12I}. Beyond the field strength itself, several additional factors may substantially influence the accretion dynamics in ULXPs, including complex, non-dipolar magnetic field configurations \citep{Israel:2016chx, Tsygankov:2018bcz}, enhanced neutrino cooling \citep{Mushtukov:2018pvq, Asthana:2023vvk}, and photon-bubble instabilities \citep{1992ApJ...388..561A, Begelman:2006ap, 2021MNRAS.508..617Z}. 

On the other hand, three primary mechanisms can contribute to neutrino emission in X-ray binaries: (i) high-energy protons accelerated within the system, such as in jets, relativistic winds, or pulsar magnetospheres. These protons undergo collisions that predominantly produce pions, which then decay into gamma rays and neutrinos \citep{Neronov:2008bw, Sahakyan:2013opa, 2023Univ....9..517K, Christiansen:2005gw,  2025A&A...701A..98D}; (ii) proton-photon interactions in the magnetosphere of the accreting pulsar \citep{Bednarek:2009vp}; (iii) electron-positron pairs of the accretion column of the bright X-ray pulsars, where pair annihilation leads to neutrino emission \citep{Mushtukov:2018pvq, 2019MNRAS.485L.131M, Asthana:2023vvk, Mushtukov:2024efo}. 

In previous studies of electron-positron pair annihilation in the accretion columns of bright X-ray pulsars and ULXPs, however, the properties of the thermal mound, as well as the structure and equation of state (EOS) of the dense matter inside pulsars, were often neglected \citep{Mushtukov:2018pvq, Asthana:2023vvk}. 
In fact, the EOS of supranuclear matter inside compact objects remains one of the most challenging problems in both microphysics and astrophysics, hopefully to be solved with the multi-messenger astronomy~\citep{2004Sci...304..536L,2019PrPNP.10903714B}.
Due to the non-perturbative nature of the fundamental strong interaction at low energies, the EOS of dense matter at nuclear densities is still poorly understood, nevertheless, strangeness should play an important role in this energy scale~\citep{2023AdPhX...837433L}.
\citet{Witten:1984rs} proposed that the true ground state of dense matter might be quark matter, composed of nearly free $u$, $d$, and $s$ quarks.
This concept suggests that compact objects resembling pulsars could be quark stars (QSs) rather than traditional neutron stars (NSs), as outlined in the MIT bag model in the asymptotic freedom regime~\citep{Alcock:1986hz}.
\citet{Xu:2003xe} proposed, however, that building units of supranuclear matter could instead be strange quark clusters (renamed ``strangeons''), rather than itinerating quarks, due to the fact that the coupling between them is still very strong~\citep{Lai:2017ney}.
These strangeons, formed by bound $u$, $d$, and $s$ quarks, represent a unique state of matter in which quarks condense in position space rather than momentum space. The term strangeon stars (SSs) was coined to describe such compact objects \citep{Xu:2016uod, Lai:2017ney, 2022MNRAS.516.6172L, 2025PhRvD.111f3033Y}.

Strangeon matter, like strange quark matter, consists of nearly equal numbers of $u$, $d$, and $s$ quarks. However, in contrast to strange quark matter, quarks in strangeon matter are localized inside strangeons due to the strong quark-quark coupling. NSs, QSs, and SSs share similarities but also differ significantly. In SSs, quarks are localized in strangeons, much like neutrons in NSs. However, unlike neutrons, a strangeon can contain more than three valence quarks, restoring light-flavour symmetry. Additionally, SSs are self-bound by the strong force, with their surface matter also composed of strangeons, similar to QSs \citep{Xu:2003xe}.

SSs can explain many key observational phenomena in astrophysics. The EOS of SSs is sufficiently stiff to account for the observed masses of pulsars \citep{Demorest:2010bx, Antoniadis:2013pzd}, while pulsar glitches may be attributed to starquakes \citep{Zhou:2014tba}. Furthermore, SSs are capable of explaining X-ray flares and bursts in magnetar candidates \citep{Xu:2006mp}, the plateau phase in gamma-ray bursts \citep{2011SCPMA..54.1541D}, the quasi-periodic oscillations in SGR
1806$-$20 \citep{Li:2023tng}, and shaping the energy budget of various power sources in the context of magnetar starquake triggering mechanisms \citep{Wang:2024opz}. Additionally, the strangeness barrier plays a crucial role in understanding Type I X-ray bursters \citep{2015ApJ...798...56L} and X-ray-dim isolated neutron stars \citep{Wang:2016nqt, Wang:2017hgt}.

In this work, we investigate the thermal mound properties within the SS framework \citep{Xu:2003xe} and estimate the neutrino luminosity relevant to ULXPs.
A notable feature of the polar mound of SSs is the sharp discontinuity in mass density, resulting from both the Coulomb~\citep{Xu:1999up, Hu:2001se, Xu:2014wxa} and the strangeness~\citep{Xu:2014wxa, 2015ApJ...798...56L, Wang:2016nqt, Wang:2017hgt} barriers.
%
Most of the total gravitational energy of accretion is stored in the matter above the surface of density discontinuity on the polar cap of an SS, whereas an NS does not exhibit such a discontinuity.
This therefore results in a lower effective heat capacity for SS and a higher one for NS, and consequently a high temperature and large neutrino emissivity of SS.
The detailed quantitative calculations in the following sections demonstrate consistency with these qualitative results.

The paper is organized as follows. In Sec. \ref{sec: model}, we calculate the temperature distribution within the accretion column and derive the thermal mound height. The luminosity associated with heat transport from the SS surface is presented in Sec. \ref{sec: Heat transport}. In Sec. \ref{sec: NE}, we estimate the neutrino energy flux based on the thermal mound properties and discuss the detectability of neutrinos from these ULXPs. Finally, we summarize and discuss our results in Sec.~\ref{sec: summary}.

\begin{table*}
    \renewcommand\arraystretch{2.0}
    \centering
    \caption{Observed properties of known ULXPs. From left to right, the table lists: source name, period, period derivative, X-ray luminosity, distance, and the corresponding reference.}
    \setlength{\tabcolsep}{0.05cm}
    {\begin{tabular}{l c c c c c c c c}
    \hline
    \hline
    Source   & $P$   & $\dot{P}$   & $L_{\rm X}$   & $d$   & Ref.
    \\[-0.5em]
             & $(\rm s) $   & $(10^{-9}\, \rm s \,s^{-1})$   & $(10^{39}\,\rm erg\, s^{-1})$   & $(\rm Mpc)$   & \\
    \hline
    M82 X-2                & 1.37   & $-0.2$   & $18$        & 3.6   & [1] \\
    NGC 5907 ULX-1         & 1.13   & $-3.8$   & $\sim 100$  & 17.1  & [2] \\
    NGC 300  ULX-1         & 16.6   & $-556$   & $4.7$       & 2     & [3] \\
    NGC 1313 X-2           & 1.5    & $-13.8$  & $\sim 20$   & 4     & [4] \\
    NGC 2403 ULX           & 18     & $-100$   & $1.2$       & 3.2   & [5] \\   
    NGC 7793 P13           & 0.417   & $-0.04$  & $\sim 10$   & 3.6   & [6] \\
    NGC 7793 ULX-4         & 0.4    & $-35$    & $3.4$       &3.9    & [7] \\
    NGC 4559 X-7           & 2.6    & $+1$     & $20$        & 7.5   & [8] \\
    Swift J0243.6$+$6124   & 9.86   & $-22$    & $2$         & 0.007 & [9] \\
    SMC X-3                & 7.8    & $-6.46$  & $1.2$       & 0.062 & [10] \\
    RX J0209.6$-$7427      & 9.29   & $-17.5$  & $1.6$       & 0.06  & [11]\\
    M51 ULX-7              & 2.8    & $-0.15$  & $4$         & 8.6   & [12]\\
    M51 ULX-8              & --    & --  & $2$         & 8.58   & [13] \\
    NGC 4559 X7            & 2.6   & $-1$ & $5$         & 7.5  & [14]\\
    \hline
    \end{tabular}}
     \begin{list}{}{}
           \item $[{1}]$ \citet{Bachetti:2014qsa}; 
          \item $[{2}]$ \citet{Israel:2016chx, Fuerst:2023mrt}; 
           \item $[{3}]$ \citet{Carpano:2018pot, Vasilopoulos:2018vqh, Walton:2018ecy};
           \item $[{4}]$ \citet{Sathyaprakash:2019stu}; 
           \item $[{5}]$ \citet{Trudolyubov:2007nh};
           \item $[{6}]$ \citet{Motch:2013zth, Furst:2016mgk, Israel:2016sxx, Fuerst:2018ybc};  
           \item $[{7}]$ \citet{Quintin:2021beb}; 
           \item $[{8}]$ \citet{Pintore:2021zbc, Pintore:2025kcn};
           \item $[{9}]$ \citet{NICER:2018evw}; 
           \item $[{10}]$ \citet{Weng:2017qrc}; 
           \item $[{11}]$ \citet{Vasilopoulos:2020jtu};
           \item $[{12}]$ \citet{RodriguezCastillo:2019esm, Vasilopoulos:2019bhi};
          \item $[{13}]$ \citet{Brightman:2018fsb, Bachetti:2021yxj};
          \item $[{14}]$ \citet{2025A&A...695A.238P}.
\end{list}
    \label{tab: Table_1}
\end{table*}

\section{Theoretical model}
\label{sec: model}

\subsection{Dynamical structure of the accretion column}
\label{sec: acc_column}

\begin{figure}
    \centering 
    \includegraphics[width=8cm]{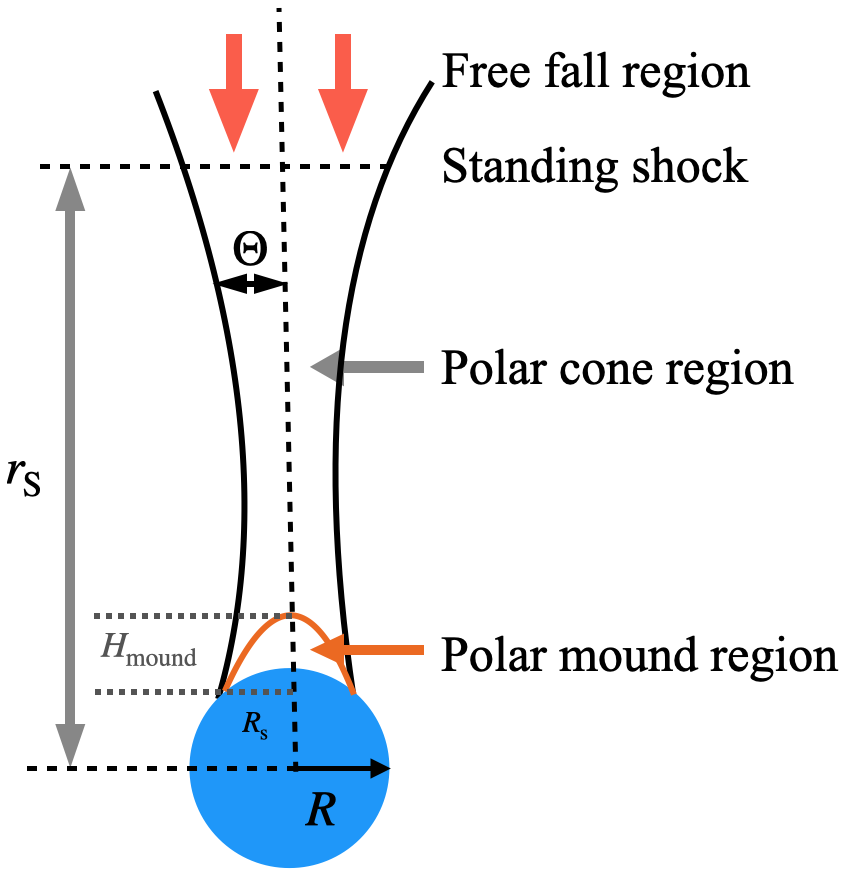}
    \caption{Schematic depiction of the magnetic polar regions with  radius $R_{\rm s} \simeq10^{5}\, \rm cm $ and height $H_{\rm mound}$ of polar mound in an X-ray pulsar. $\Theta$ is the opening angle of the magnetic funnel. This diagram is adapted from \citet{2020PASJ...72...12I}.}
    \label{fig: diagram}
\end{figure}

In this subsection, we discuss the structure of the accretion column and calculate key quantities, including the radiation energy, mass density, and velocity of the mass flow within the column. 

Radiation pressure plays a crucial role in determining the dynamical structure of accretion flows in high-luminosity X-ray pulsars. At sufficiently high accretion rates, a shock forms above the stellar surface \citep{1975PASJ...27..311I, 1976MNRAS.175..395B, 2020PASJ...72...12I}. This shock must be radiative in nature to effectively dissipate the kinetic energy of the infalling matter \citep{1998ApJ...498..790B}. Models of such accretion columns are essential for interpreting observed radiation luminosities \citep{Mushtukov:2015zea}, pulse profiles \citep{Klochkov:2008jj, 2012A&A...544A.123B}, continuum spectra \citep{Becker:2006tt}, cyclotron line spectra \citep{Staubert:2018xgw}, and polarization observations of high-luminosity X-ray pulsars in high-mass X-ray binaries. Moreover, the accretion column exhibit quasi-periodic oscillations using the magnetohydrodynamics simulations \citep{2022MNRAS.515.4371Z, 2023MNRAS.520.1421Z, 2025MNRAS.540.3934Z} 

A schematic diagram of the magnetic polar regions is shown in Fig. \ref{fig: diagram}, which includes the free-fall region, the polar cone region, and the polar mound region. At the free-fall region, matter can be approximated to flow with the free-fall velocity on the upstream side of the shock. The kinetic energy of the inflowing matter is converted into thermal energy through the shock in the polar cone region, where radiation is emitted laterally in a fan beam pattern. At the base of the polar cone, a thermal mound forms on the stellar surface, and radiation is emitted radially upward in a pencil beam.

In highly magnetized accreting pulsars, the plasma is funneled by the strong dipolar magnetic field toward the magnetic polar caps, forming narrow accretion columns above the stellar surface. In these regions, the inflowing matter is nearly constrained to move along magnetic field lines, and the lateral motion can be neglected compared to the radial inflow. Assuming one-dimensional, steady-state accretion, the structure of the accretion column is determined by the conservation of mass, momentum, and energy  \citep{1975PASJ...27..311I, 1976MNRAS.175..395B, 1998ApJ...498..790B}. 

Following the variational derivation of  \citet{1975PASJ...27..311I}, the dynamical equations describing the vertical structure of the magnetic polar regions can be written as \citep[see also][]{2020PASJ...72...12I}, 
\begin{align} 
\frac{{\rm d}\varepsilon}{{\rm d}r} & = \frac{3}{4} ( \frac{\varepsilon}{r_{\rm D}} - \frac{GM}{r^{2}} )  \label{eq: dyna_1} \,, \\ 
\frac{{\rm d}\rho}{{\rm d}r} & = -\frac{3\rho}{4\varepsilon} ( \frac{\varepsilon}{r_{\rm D}} + 3  \frac{GM}{r^{2}} )  
\label{eq: dyna_2}\,.
\end{align}
Here, $\varepsilon$ is the energy density of the accreting plasma, $\rho$ is the mass density. 
The effective photon diffusion length $r_{\rm D}$ characterizes the scale over which the inflowing kinetic energy is converted into radiation via radiative drag and is given by
\begin{equation}\label{eq: r_D}
r_{\rm D} \equiv \frac{ 3 \kappa_{\rm Th} \dot{M}}{8 \pi c} = 1.4 \times 10^{5} \left(\frac{\dot{M}}{10^{17} \rm{g \; s}^{-1}} \right) \; \rm{cm} \,,
\end{equation}
where $\kappa_{\rm Th}$ is the opacity of the Thomoson scattering in the tangential direction of the polar cone, $\dot{M}$ is the mass accretion rate onto the stellar surface. The quantity $r_{\rm D}$ represents the distance that the infalling matter advances within the time it takes for a photon to diffuse from the central axis to the surface of the cone.
Eq. (\ref{eq: dyna_1}) describes the evolution of the energy density, balancing gravitational acceleration  $( GM/ r^{2})$ and radiative dissipation. Eq.  (\ref{eq: dyna_2}) governs the density stratification, where the compressive effects of gravity compete with radiative pressure support. Together, these equations determine the vertical structure of the accretion column and provide the foundation for understanding the formation of the thermal mound and the associated energy transport in the polar regions.

We adopt the same boundary conditions as in \citet{2020PASJ...72...12I}. At the top boundary of the polar cone, we set the mass density to
\begin{equation}\label{eq: boundary conditions_1}
  \rho_{\rm top} = 7 \rho_{\rm F}(r_{\rm S}) \,,
  \end{equation}
where $\rho_{\rm F}$ is the matter density in the free-fall region at radius $r_{\rm S}$. The radiation energy density at the top boundary is given by
\begin{equation}\label{eq: boundary conditions_2}
  \varepsilon_{\rm top} = \frac{3}{4}\frac{GM}{r_{\rm S}} \,,
\end{equation}
which reflects the gravitational energy conversion near the shock front. Since both $\rho_{\rm top} $ and $\varepsilon_{\rm top}$ depend on the location of the top boundary $r_{\rm S}$, we numerically integrate the dynamical equations inward from an assumed value of $r_{\rm S}$ toward the stellar surface, decreasing $r$ until the bottom boundary condition is satisfied. At the stellar surface, the bottom boundary is determined by the balance between the gas pressure and magnetic pressure 
\begin{equation}\label{eq: boundary conditions_3}
  P_{*} = \frac{\rho_{*} \varepsilon_{*}}{3} =\frac{ B^2_{*}}{8 \pi}\,,
\end{equation}
where the subscript $*$ denotes the physical quantities evaluated at the stellar surface. The location of the top boundary $r_{\rm S}$ is thus iteratively adjusted to ensure consistency with this bottom boundary condition.

\begin{figure*}
    \centering 
    \includegraphics[width=12cm]{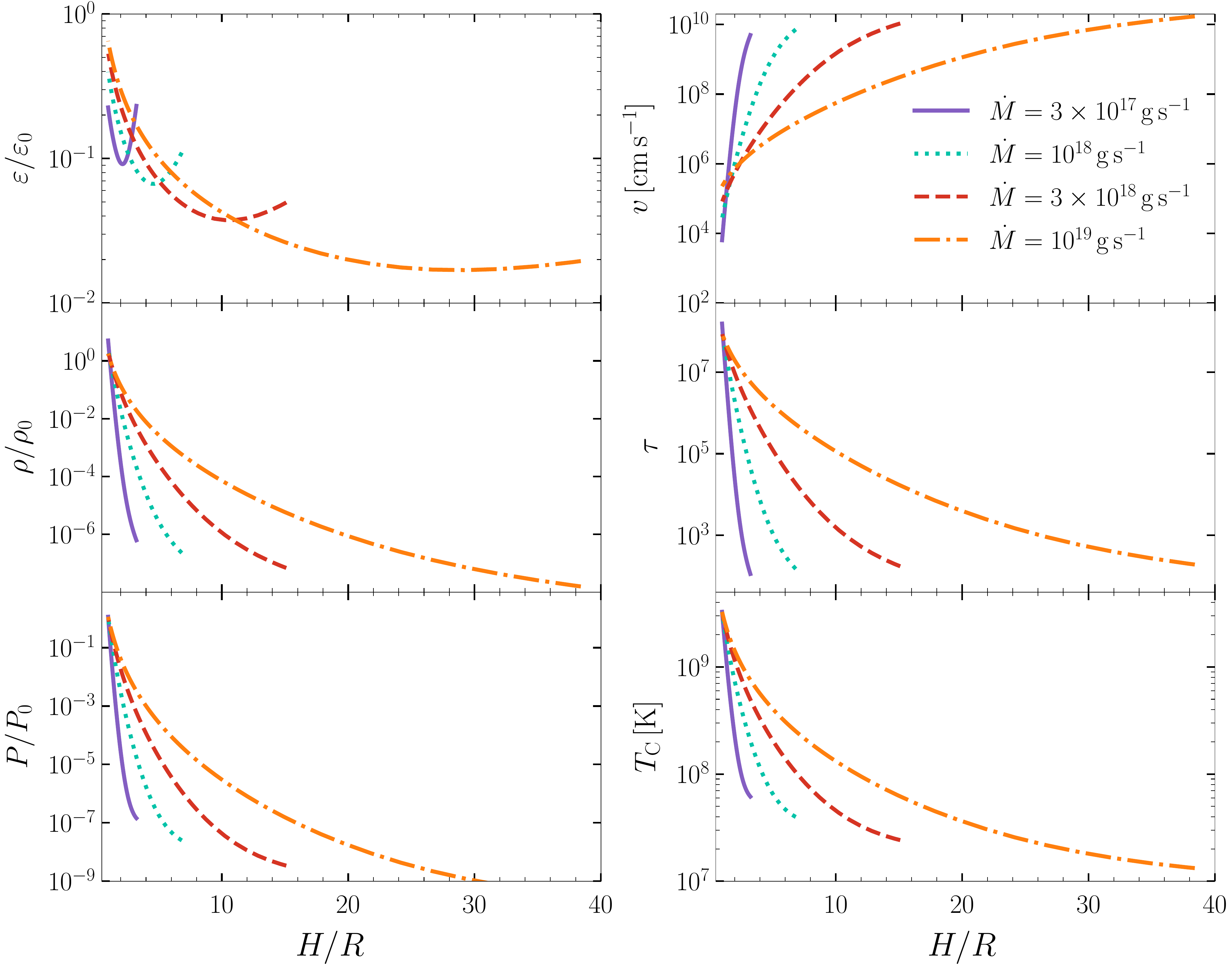}
    \caption{Profiles of physical quantities within the magnetic polar regions as functions of height, with different colors indicating different mass accretion rates. We adopt a stellar model with $M = 1.4 \, M_{\odot}$ and $R = 10^{6}\, \rm cm$. The first column presents the radiation energy density, mass density, and pressure, while the second column shows the corresponding velocity, optical depth, and temperature profiles.  Normalization parameters are defined as $\varepsilon_{0} = GM/R$, $\rho_{0}=3 B^2 R/ (8 \pi G M) \sim 10^{2}\, \rm g\,cm^{-3}$, and $P_0 =B^2/ (8\pi) $, assuming a surface magnetic field strength of $B = 10^{12}\, \rm G $. The characteristic temperature, $T_{\rm C}$, is calculated via $T_{\rm C} = (\rho \varepsilon/a)^{1/4}$, where $a$ is the radiation constant. The height of the polar mound $H_{\rm mound} \approx  0.8 \, \rm km$.}
    \label{fig: column_SS}
\end{figure*}

In Fig. \ref{fig: column_SS}, we show the radiation energy, mass density, pressure,  velocity, optical depth, and temperature as functions of the column height for different mass accretion rates.
The optical depth  $\tau $ of the Thomson scattering in the tangential direction of the polar cone is given by \citep{2020PASJ...72...12I}
\begin{equation}\label{eq: tau}
  \tau  = \kappa_{\rm Th} \rho r \Theta_{0} \,.
\end{equation}
Here, the opening angle $\Theta_{0}$ of the magnetic polar region can be written as \citep{2020PASJ...72...12I}
\begin{align}
\Theta_{0} = & \,
4 \times 10^{-2} \, \left(\frac{R}{10^6\, \rm cm} \right)^{1/2} \left(\frac{\dot{M}}{10^{17}\, \rm g\,s^{-1}} \right)^{1/7} \left(\frac{M}{M_{\odot}} \right)^{1/14}
\nonumber
\\
& \,
\left(\frac{\mu_{\rm M}}{10^{30}\, \rm G\, cm^3} \right)^{-2/7} \,,
\label{eq: Theta}
\end{align}
where $\mu_{\rm M}$ is the magnetic dipole moment of the pulsar. It can be observed that an extremely high mass accretion rate results in a taller accretion column. Moreover, the pressure and temperature at the base of the column remain independent of the accretion rate because the magnetic pressure counterbalances the radiation pressure.

In particular, the temperature in the polar mound exceeds $10^{9}\, \rm K$, leading to energy losses through neutrino emission \citep{1967ApJ...150..979B, 1994A&AT....4..283K}. The primary mechanism for neutrino production is the annihilation of electron-positron pairs, $e^{-} + e^{+}\rightarrow \nu + \bar{\nu}$ \citep{1967ApJ...150..979B}.  At high temperature regime the pair annihilation remains the dominant process for neutrino production with the corresponding energy losses of $Q_{\nu} \approx 4 \times 10^{24}\, \left( \frac{T}{10^{10}\, \rm K} \right)^{9}\,  {\rm erg\, cm^{-3}\, s^{-1}}$.

\subsection{Calculation of the polar mound height}
\label{sec: mound height}

A key question in ULXPs is the region capable of effectively radiating neutrinos. \citet{Mushtukov:2018pvq} proposed that strong advection at extremely high mass accretion rates leads to the formation of a zone in the central region of the accretion column, where the released energy is carried downward and emitted as neutrinos. In their study, the height of the neutrino emission region is determined by photon advection. However, their model did not consider the thermal mound properties, nor did it address how the structure and EOS of the pulsar could significantly influence the thermal mound characteristics and the efficiency of neutrino emission.

In this work, we investigate the thermal mound properties based on SS model \citep{Xu:2003xe}, taking into account  both the Coulomb barrier \citep{Xu:1999up, Hu:2001se, Xu:2014wxa} and the strangeness barrier \citep{Xu:2014wxa, 2015ApJ...798...56L, Wang:2016nqt, Wang:2017hgt}. The inflowing matter penetrates the Coulomb barrier and follows the mass conservation equation,
\begin{equation}\label{eq: mound_1}
\dot{M}(R_{M}) = \tau_{\rm cou} \frac{{\rm d} N_{n}}{{\rm d} t} m_{p} \,,
\end{equation}
where $\dot{M}(R_{M})$ is the mass accretion rate at the magnetospheric radius, $m_{p}$ is the proton mass, and $\tau_{\rm cou}$ is the penetration probability, given by
\begin{equation}
\tau_{\rm cou} = \exp \left[-\left(\frac{E_{\rm G}}{E}\right)^{1/2}\right] \,,
\end{equation}
with $E_{\rm G} = 0.49\, \rm MeV$ as the Gamow energy \citep{2003paas.book.....P}, and $E = k_{B} T_{\rm base}$, where $k_{B}$ is the Boltzmann constant and $T_{\rm base}$ is the base temperature of the accretion column, which depends on the mass accretion rate (see Table \ref{tab: Table_5907}).
The penetration time of the nucleons is approximately $ t = 0.01 c/(1/n)^{1/3} \approx 1 \times 10^{17}\, \rm s $. The number of nucleons attempting to penetrate is $N_{n} = \lambda n S h$, where $n$ is the number density, $h \approx (1/n)^{1/3}$. $S =\pi R_{\rm S}^2 \approx 10^{10} \, \rm cm^2$ is the cross-sectional area of the accretion column base. However, the size of the cross-sectional area of the accretion column base remains a topic of ongoing debate \citep{1973ApJ...184..271L, 1976ApJ...207..914A, 1980ApJ...235.1016A, 1977ApJ...215..897E, 1976Natur.262..356E, 1984ApJ...278..326E, 1977ApJ...216..838M, 1977ApJ...214..261M, 1977ApJ...213..836M, 1977ApJ...217..578G, 1977ApJ...214..550P, 1977ApJ...216..827P, 1977ApJ...218..783P, 1998ApJ...498..790B, 2007ApJ...654..435B, 2022ApJ...939...67B, 2017ApJ...835..129W, 2017ApJ...835..130W}. In the works of \citet{2007ApJ...654..435B} and \citet{2022ApJ...939...67B}, the shape of the X-ray continuum in accretion-powered pulsars has been successfully fitted using the accretion column model. The cross-sectional area of the accretion column in their models is smaller than that in \citet{Asthana:2023vvk}. Therefore, we set the parameter  $\lambda = 0.01$ to modify the size of the cross-sectional area. 
Our results indicate that the Coulomb barrier in strangeon stars does not effectively prevent the inflowing matter from penetrating the star.


Furthermore, the falling matter may significantly penetrate the strangeness barrier, most of the falling matter would be bounced back along the magnetic field lines  because nonstrange matter cannot become part of strange matter unless it is converted to strangeons via weak interaction \citep{Xu:2014wxa, Wang:2016nqt}. This process is key to the formation of the thermal mound.

To estimate the thermal mound height, we first calculate the number density of nonstrange ions accumulated at the base of the polar cone. As discussed earlier, the conversion between $u$, $d$, and $s$ quarks is governed by weak interactions. Motivated by the intrinsic weak decay lifetime of strange baryons such as the $\Lambda$ hyperon, $\tau _{\Lambda} \approx 2.6 \times 10^{-10}\, \rm s$, we adopt a typical weak interaction timescale of $t^{*}_{\rm weak} = 10^{-10}\, \rm s $. Under this condition, the probability for incident ions to undergo flavor conversion into strangeons is estimated to be  $\eta \sim 10^{-13}$, assuming an inflow velocity of $v_{i} \sim 0.01 c$.
The number density at the base of the polar mound is related to the mass accretion rate via
\begin{equation}\label{eq: mound_2}
\eta\,  T_{N}\,  n_{\rm base}\, S h = \dot{M}(R_{M})/m_{p}  \,,
\end{equation}
where the number of collisions $T_{N}= [(n_{\rm base})^{-1/3}/v_{i}]^{-1}$. Using Eqs. (\ref{eq: mound_1}) and (\ref{eq: mound_2}), we obtain that the number density and mass density of the nonstrange normal matter at the base of the polar mound are $ n_{\rm base} \sim 10^{35}\, \rm cm^{-3} $, $ \rho_{\rm base} \sim 10^{11}\, \rm g \, cm^{-3} $, respectively.  

In a one-dimensional model with the area of accretion column base, we can express the pressure contained in the accretion column as a function of $r$ using
\begin{equation}\label{eq: eq_1}
\frac{{\rm d} P}{{\rm d} r} = -\rho \frac{G M}{(R+r)^2} \,,
\end{equation}
which is related to EOS. We consider contributions from gas pressure, radiation pressure, and degenerate pressure, as described in \citet{1983ApJ...267..315P}. Using Eq. (\ref{eq: eq_1}) in conjunction with EOS, we can obtain the height of the thermal mound. 
We note that magnetic pressure, $P_{\rm mag} = B^2/(8\pi)$, although important for confining the column globally, varies negligibly over the small height of the polar mound ($H_{\rm mound} \approx 1\,\rm km$). Its contribution to the pressure gradient is therefore very small and can be safely neglected in estimating the mound height.

\begin{figure}
    \centering 
    \includegraphics[width=8cm]{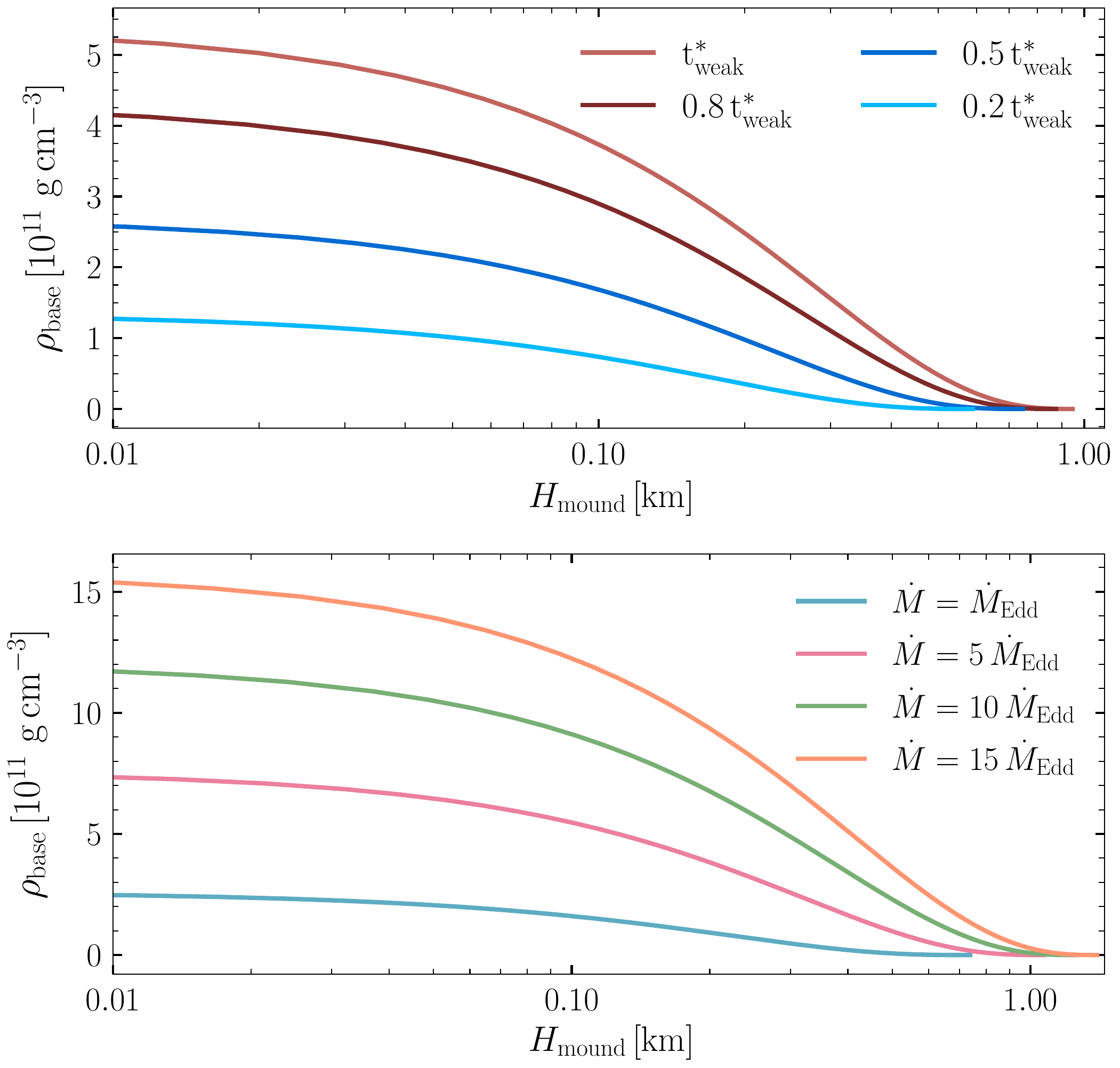}
    \caption{Mass density of the thermal mound  as a function of thermal mound height for different weak interaction timescales with a fixed mass accretion rate  $\dot{M}=  \dot{M}_{\rm Edd}$ ({\it upper panel}), and for different mass accretion rates with a fixed weak interaction timescale $ t^{*}_{\rm weak} = 10^{-10}\, \rm s $  ({\it lower panel}). We set magnetic field strength $B = 10^{12}\, \rm G $. }
    \label{fig: mound}
\end{figure}

The upper panel of Fig. \ref{fig: mound} shows the mass density of the thermal mound as a function of thermal mound height for different weak interaction timescales, with a fixed mass accretion rate $\dot{M} = \dot{M}_{\rm Edd}$. The thermal mound height $H_{\rm mound}$ ranges from $0.7$ to $0.95\, \rm km$. The lower panel of Fig. \ref{fig: mound} shows the mass density as a function of thermal mound height for different mass accretion rates. As the mass accretion rate increases, the thermal mound height becomes larger.

The density at the base of the thermal mound in SSs is approximately $10^{11}\, \rm g \, cm^{-3} $, while the surface density of strangeon matter exceeds $10^{14}\, \rm g \, cm^{-3}$, resulting in a discontinuity  in mass density at the surface. This discontinuity reduces the number of particles in the polar mound region of an SS, lowering its effective heat capacity. With thermal energy supplied by gravitational potential energy, the reduced heat capacity leads to a sharply elevated temperature, which, together with the high density, promotes efficient neutrino production.
In contrast, the surface density of an NS is continuous and generally lower than $10^{11}\, \rm g \, cm^{-3} $ \citep{2012MNRAS.420..720M, 2017JApA...38...48M}, and the thermal mound height in NSs is significantly smaller than in SSs \citep{2013MNRAS.430.1976M, 2013MNRAS.435..718M}. In an NS, the same thermal energy is distributed over more matter, resulting in a higher heat capacity and a much lower polar mound temperature, rendering neutrino emission inefficient. These contrasts highlight the distinctive thermal and neutrino emission properties of SSs relative to NSs.

\section{Heat transport from strangeon star surface}
\label{sec: Heat transport}

In this section, we discuss the heat energy transport from SS surface.  According to our results, the temperature at the base of the polar mound exceeds $10^9\, \rm K$, and this thermal energy can be transported along the SS surface to other regions.  The heat flux is related to the temperature gradient through Fourier’s law, ${\bf F} = - \kappa \nabla T $, where the thermal conductivity $\kappa$ is primarily determined by electron-phonon interactions. The typical value is  $1.2 \times 10^{20} \, \rm erg\,K^{-1}\, cm^{-1}\, s^{-1} $  \citep{1976ApJ...206..218F, 2009arXiv0905.3818Y}.

 In this work, we adopt a one-dimensional approximation for the temperature gradient, $\nabla T \simeq (T_{\rm base} - T_p) / R_{\rm s}$, where $T_{\rm base}$ is the base temperature of the accretion column (see Table \ref{tab: Table_5907}),  $T_p \simeq 2-3 \, \rm keV$ \citep{Middleton:2014kva, 2025MNRAS.536..340K}. Assuming that the high-temperature region corresponds to the base of the accretion column, the luminosity associated with heat transport can be expressed as
\begin{equation} 
L_{\rm heat} = 4 \pi F S \approx 1.2 \times 10^{36}\, \rm erg\, s^{-1}\,.
\end{equation}
It should be noted that $L_{\rm heat}$ is independent of the mass accretion rate. The bottom of the thermal mound is connected to the SS surface, allowing energy to be transported as heat. Meanwhile, the thermal mound loses energy through neutrino emission, and the outer regions of the accretion column also radiate energy via photons. In the following section, we discuss the neutrino and photon luminosities separately and compare the estimated flux levels with the neutrino background and the sensitivities of current neutrino telescopes.

\section{Neutrino flux and detection}
\label{sec: NE}

Relying on a model of the thermal mound with high temperature of accreting SS, we calculate the neutrino luminosity, and estimate the neutrino energy flues for different ULXPs. The total accretion luminosity can be written as
\begin{equation}\label{eq: L_total}
L_{\rm tot} = \frac{G M \dot{M}}{R} = L_{\rm \nu} 
+ L_{\rm ph}  + L_{\rm heat} \,,
\end{equation}
where $L_{\rm ph}$ and $L_{\rm \nu}$ are the neutrino and photon luminosity, respectively. Following the methodology of  \citet{Mushtukov:2018pvq, Asthana:2023vvk},  the neutrino luminosity is given by 
\begin{equation}\label{eq: L_neu}
 L_{\rm \nu} \approx S \int^{H_{\rm mound}}_{0} Q_{\rm acc}(H_{\rm mound}) dH_{\rm mound} \,,
\end{equation}
where $H_{\rm mound}$ is the height of the thermal mound, which is dependent on the weak interaction timescale and mass accretion rate (see Fig. \ref{fig: mound}). The accretion heating rate defined by variations of the gravitational potential energy is given by 
\begin{equation}\label{eq: Q_acc}
 Q_{\rm acc}(H_{\rm mound}) = \frac{\dot{M}}{2 S} \left[\frac{G M}{(R + H_{\rm mound})^2} \right]  \,.
\end{equation}
Here, we primary consider the the gravitational potential energy, and neglect the effect of the kinetic energy and the radiation energy for the radiation-dominated shock at the top of accretion column. 

\begin{figure}
    \centering 
    \includegraphics[width=8cm]{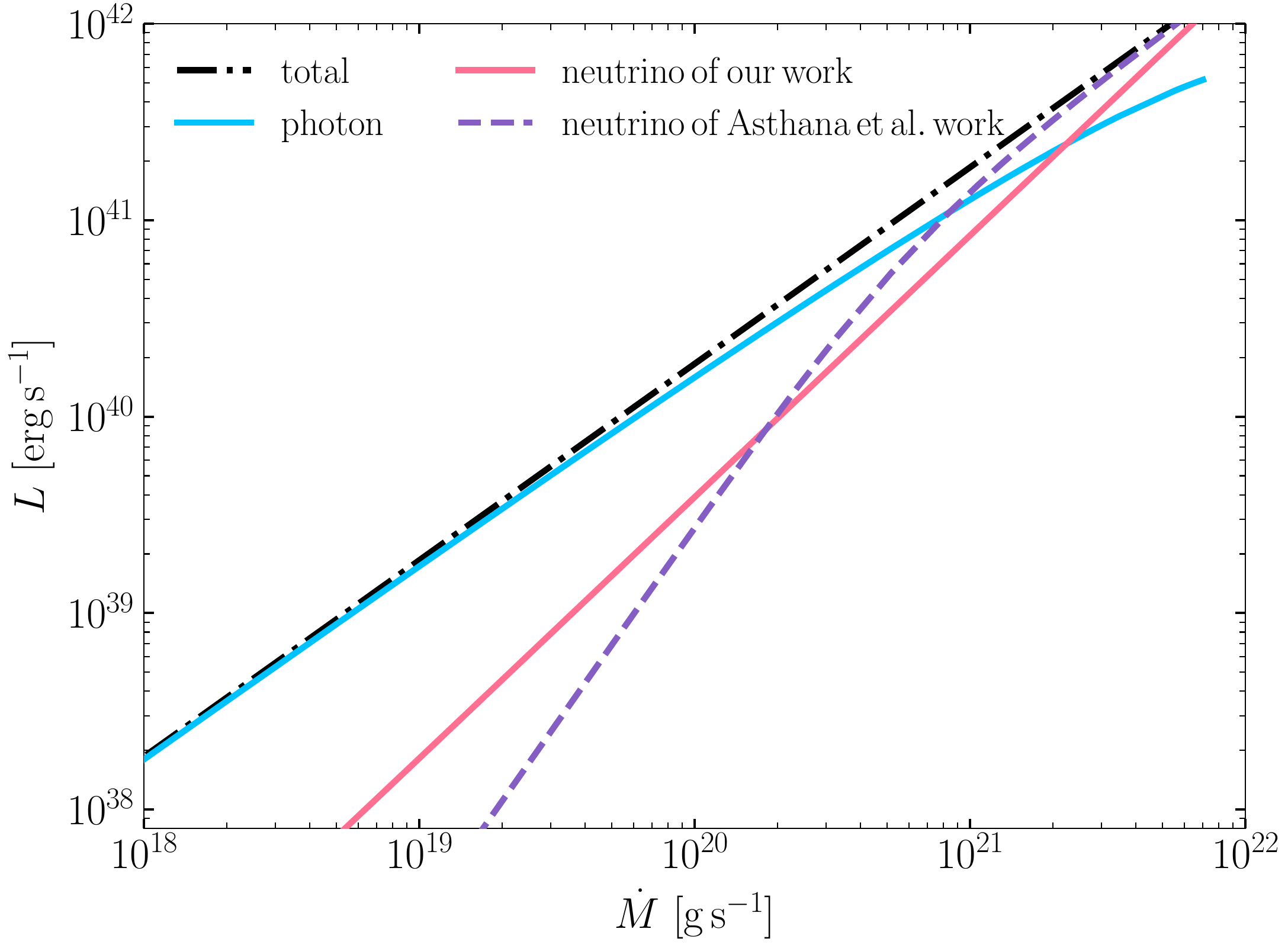}
    \caption{The neutrino luminosity of our work (solid line), neutrino luminosity of \citet{Asthana:2023vvk} (dashed line), and the photon luminosity
    as functions of the  mass accretion rate. The dash-dotted line is total luminosity is the dash-dotted line. The horizontal dash lines are show the observed luminosity for some ULXPs. We calculate the neutrino luminosity with $t = t^{*}_{\rm weak}$ and the magnetic field strength $B = 10^{12}\, \rm G $ in our work.}
    \label{fig: luminosity}
\end{figure}

In Fig. \ref{fig: luminosity}, we show the total, neutrino, and photon luminosity as functions of the mass accretion rate.  We observe that the results are similar in the following scenarios. At relatively low mass accretion rates ($< 10^{20}\, \rm g\,s^{-1}$), photons can escape from the accretion column, and the neutrino luminosity remains negligible, with the total luminosity being dominated by photons. However, at higher mass accretion rates ($ >10^{21}\, \rm g\,s^{-1}$), a large region forms where photons are trapped, and energy is transported to regions cooled by neutrinos, leading to a total luminosity that exceeds the photon luminosity. Furthermore, as the mass accretion rate increases, most of the energy is converted into neutrinos, causing the photon luminosity growth to slow. The photon luminosity may approach an upper limit of around  $10^{41}\, \rm erg\, s^{-1}$.

For comparison with the NS model, we also compute the neutrino luminosity from the accretion column using Eq. (1) of \citet{Asthana:2023vvk} (dashed lines in Fig. \ref{fig: luminosity}). At relatively low accretion rates  ($< 10^{20}\, \rm g\,s^{-1}$), the SS model predicts moderately higher neutrino fluxes for Swift J0243.6$+$6124, RX J0209.6$-$7427, and SMC X-3, exceeding the results of \citet{Asthana:2023vvk} by factors of $\sim 2-5$. In contrast, for ULXPs, the SS model yields neutrino fluxes that are comparable to or lower than those of the NS model, with typical reductions of $\lesssim 2$ and up to a factor of $\sim 5$ for extreme sources such as NGC 5907 ULX-1. This comparison suggests that, within the SS framework, nearby Be X-ray binaries (e.g. Swift J0243.6$+$6124 and RX J0209.6$-$7427) are more promising neutrino sources than ULXPs.

The possibility of neutrino detection on Earth depends not only on the neutrino flux but also on the species composition and spectral distribution of the neutrinos. The dominant process of neutrino emission from the annihilation of electron-positron pairs. 

We neglect neutrino oscillations in the vicinity of the X-ray source due to the significant uncertainty in the geometry of the accretion column and its small extent. The spectra of different neutrino flavors are assumed to be the same, as the kinematics of the annihilation process are identical (see, for example, \citet{Asthana:2023vvk} for a detailed discussion). 

The photon luminosity of the sources, expected neutrino luminosity, distance to the source, and corresponding neutrino energy flux at Earth from ULXPs are summarized in Table \ref{tab: Table_2}. 
It is evident that the inferred neutrino luminosity of Swift J0243.6$+$6124 is significantly higher than that of NGC 5907 ULX-1, primarily due to the large difference in their distances. Swift J0243.6$+$6124 is located at only $0.007$ Mpc, while NGC 5907 ULX-1 lies at $17.1$ Mpc. Since the observed neutrino flux decreases with the square of the distance, this naturally results in a discrepancy of nearly four orders of magnitude. Except for Swift J0243.6$+$6124, all detected ULXPs are extragalactic, leading to substantial attenuation of their signals due to their large distances.
We discuss the possibility of neutrino detection from Swift J0243.6$+$6124, for which the neutrino flux at Earth is expected to be the highest. For various ULX Pulsar sources, the proposed model encompasses a broad range of neutron star surface magnetic field strengths, from $10^{11}\,\rm G$ to $10^{14}\, \rm G$ \citep{Furst:2016mgk, Fuerst:2018ybc, Fuerst:2023mrt, King:2023nft, 2025AN....34640106F}. In Table \ref{tab: Table_5907}, we present the effects of the magnetic field on the neutrino luminosity and corresponding neutrino energy flux for NGC 5907 ULX-1. The neutrino luminosity and corresponding neutrino energy flux change with increasing magnetic field strength, and we hope that other sources will exhibit similar properties.




To estimate an upper limit on the neutrino contribution from the broader ULXP population, we adopt a simple scaling approach. Current observational constraints suggest that there are approximately $\sim 7-30$ ULXPs in the Milky Way \citep{Shao:2015eha, Gao:2022fsg, Li:2024njw} and $\sim 384-629$ ULX candidates in nearby galaxies \citep{Earnshaw:2018krz, Kovlakas:2020wfu}. Under the simplifying assumption that these systems share similar intrinsic properties, we scale the neutrino flux of Swift J0243.6$+$6124 by a factor of  $\sim 7-30$ to represent the Galactic contribution, and the flux of NGC 5907 ULX-1 by a factor of $\sim 384-629$ to represent the contribution from nearby galaxies. We emphasize that this procedure does not constitute a population-synthesis calculation, but rather provides an optimistic upper-limit estimate of the total neutrino flux. Recent LHAASO results indicating  $\sim 10$ Galactic ULXs are consistent with this estimate within current uncertainties \citep{2024arXiv241008988L, Peretti:2024ecg, 2025arXiv251001369K}.

\begin{figure}
    \centering 
    \includegraphics[width=8cm]{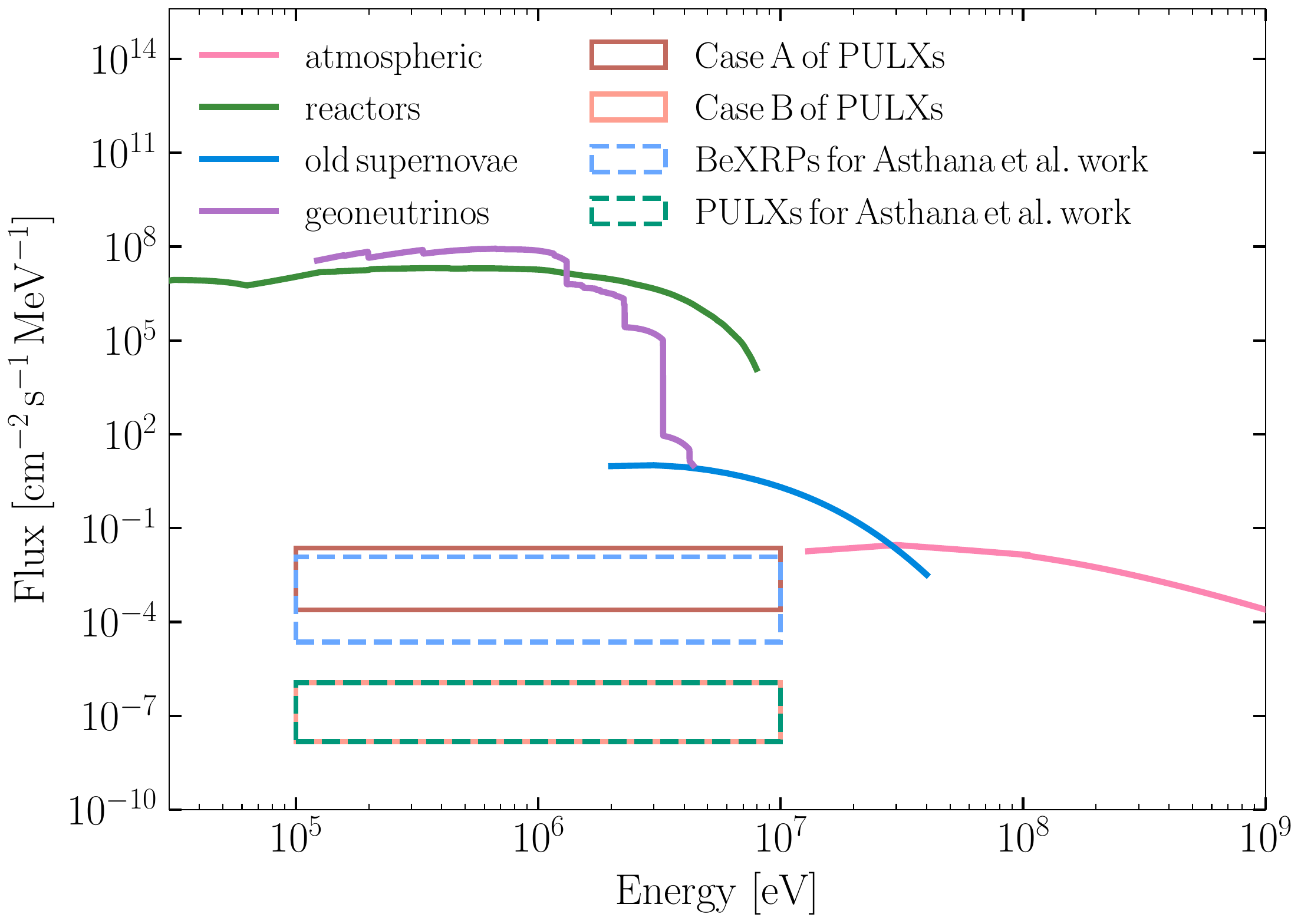}
    \caption{The energy spectrum of neutrino background at Earth in the energy range from $10^4 \, \rm eV$ to $10^{12}\, \rm eV$. The background component of the neutrino flux includes contributions from atmospheric neutrinos, nuclear reactors, old supernovae, and geoneutrinos. These data were taken from \citet{Vitagliano:2019yzm}. The open solid rectangles show the neutrino fluxes predicted by the SS model in the MeV energy band. For PULXs, Case A corresponds to Swift J0243.6$+$6124 and RX J0209.6$-$7427, while Case B includes NGC 5907 ULX-1 and M51 ULX-8. The open dashed rectangles represent the neutrino fluxes of Be X-ray pulsars and ULX pulsars, respectively, with values taken from \citet{Asthana:2023vvk}.}
    \label{fig: spectrum}
\end{figure}

In Fig. \ref{fig: spectrum}, we present the predicted neutrino background spectrum at Earth, covering the energy range from $10^4 \, \rm eV$ to $10^{12} \, \rm eV$. Neutrinos generated via electron–positron pair annihilation primarily occupy the MeV band and are initially dominated by electron neutrinos and antineutrinos. Therefore, we adopt an energy range of $0.1$–$10 \, \rm MeV$, which is appropriate for neutrinos produced via pair annihilation.
The open solid rectangles denote the neutrino fluxes predicted by the SS model. For PULXs, Case A corresponds to Swift J0243.6$+$6124 and RX J0209.6$-$7427, while Case B includes NGC 5907 ULX-1 and M51 ULX-8. The corresponding neutrino fluxes are listed in Table~\ref{tab: Table_2}. The open dashed rectangles represent the neutrino fluxes of Be X-ray pulsars and ULX pulsars, respectively. Compared with the NS model \citep{Asthana:2023vvk}, we find that the neutrino energy fluxes predicted by the SS model are broadly similar to those of the NS model.

The predicted neutrino flux spectra at Earth from the two ULXPs (Swift J0243.6$+$6124 and NGC 5907 ULX-1) are shown in  Fig. \ref{fig: spectrum}. The focus on these two systems is motivated by their distinct observational advantages. As discussed earlier, the observable neutrino flux decreases with the square of the distance. Swift J0243.6$+$6124 is by far the closest ULXP and therefore represents the most promising target for potential neutrino detection. Conversely, the neutrino flux is positively correlated with the source luminosity, and NGC 5907 ULX-1 is one of the brightest ULXs known, with an X-ray luminosity reaching $ 500\, L_{\rm Edd}$. These considerations make the two sources representative of the most favorable detection scenarios among currently confirmed ULXPs.

Even if additional nearby sources similar to Swift J0243.6$+$6124 exist but are beamed away from Earth, their neutrino emission would not be observable. Therefore, the total neutrino flux from Galactic ULXPs remains low compared with the diffuse background, and no significant detection is expected with current neutrino observatories. This highlights Swift J0243.6$+$6124 as the most promising individual target for neutrino detection.



\begin{table*}
    \renewcommand\arraystretch{2.0}
    \centering
    \caption{The photon luminosity of the objects $L_{\rm ph}$, expected neutrino luminosity $L_{\nu}$, distance to the source, and corresponding neutrino energy flux $F_{\nu}$ at Earth from ULXPs. We set magnetic field strength $B = 10^{12}\, \rm G $.}
    \setlength{\tabcolsep}{0.05cm}
    {\begin{tabular}{l c c c c c c c}
    \hline
    \hline
    Source   & $L_{\rm ph}$   & $L_{\nu}$   & $d$   &  $F_{\nu}$  
    \\[-0.5em]
             & $(10^{39}\,\rm erg\, s^{-1})$   & $(10^{39}\,\rm erg\, s^{-1})$   & $(\rm Mpc)$   & $(\rm 10^{-6}\,cm^{-2}\,s^{-1}\, MeV^{-1})$   & \\
    \hline
    M82 X-2                & $18$       & $4.5$      & 3.6     & $2.06$ \\
    NGC 5907 ULX-1 & $100$     & $64$       & 17.1   & $1.14$  \\
    NGC 300  ULX-1  & $4.7$      & $0.58$    & 2        & $0.75$  \\
    NGC 1313 X-2       & $20$      & $8.3$      & 4         & $2.71$  \\
    NGC 2403 ULX     & $1.2$     & $0.18$    &3.2       & $0.09$ \\   
    NGC 7793 P13       & $10$      & $2.4$      & 3.6      & $0.96$ \\
    NGC 7793 ULX-4  & $3.4$     & $0.58$    & 3.9      & $0.20$\\
    NGC 4559 X-7       & $20$       & $8.3$      & 7.5     & $0.77$ \\
    Swift J0243.6$+$6124  & $2$  & $0.22$    & 0.007  &$23475$ \\
    SMC X-3                & $1.2$      & $0.18$    & 0.062  &$244$  \\
    RX J0209.6$-$7427 & $1.6$    & $0.17$    & 0.06    &$246$ \\
    M51 ULX-7             &  $4$      &$0.58$     & 8.6       &$0.04$   \\
    M51 ULX-8             & $2$       & $0.22$    & 8.58      &$0.015$  \\
    \hline
    \end{tabular}}
 \label{tab: Table_2}
\end{table*}

\begin{table*}
    \centering
    \caption{The base temperature of accretion column $T_{\rm base}$, expected neutrino luminosity $L_{\nu}$,  distance to the source, and corresponding neutrino energy flux $F_{\nu}$ of the brightest source, NGC 5907 ULX-1 with the differents magnetic fields.}
    \renewcommand\arraystretch{1.25}
    \begin{tabular}{c c c c c }
    \hline
    \hline
    B  &  $T_{\rm base}$ &  $L_{\nu}$ & $d$  & $F_{\nu}$  \\
    (G) & $(10^{9}\,\rm K)$  & $(10^{39}\,\rm erg\, s^{-1})$  & $(\rm Mpc)$ &$(\rm 10^{-6}\,cm^{-2}\,s^{-1}\, MeV^{-1})$  \\
     \hline
    $10^{12}$  & 3         &   64     &  $17.1$     &  1.14   \\
    $10^{13}$  & 10      &   139       &  $17.1$    &   2.64 \\
    $10^{14}$  & 32    &     216      &   $17.1$     &   4.12  \\
    \hline 
    \end{tabular}
    \label{tab: Table_5907}
\end{table*}

\section{Summary and discussion}
\label{sec: summary}

In this paper, we have investigated the structure and emission properties of accretion columns in ULXPs, focusing on the role of the thermal mound that forms at the base of the column on a SS. Using a one-dimensional approximation for the temperature gradient and a detailed treatment of the Coulomb and strangeness barriers, we calculated the height and physical properties of the thermal mound, including mass density, pressure, and temperature. Our results indicate that the thermal mound can reach heights of $ 0.7 \sim 0.95 \, \rm km $, with base temperatures exceeding $10^9\, \rm K$, sufficient to produce significant neutrino emission via electron-positron annihilation.

We find that at relatively low mass accretion rates  ($< 10^{20}\, \rm g\,s^{-1}$), photons can escape freely from the accretion column, and the total luminosity is dominated by photon emission, while neutrino luminosity remains negligible. At higher mass accretion rates ($ >10^{21}\, \rm g\,s^{-1}$), photon trapping becomes significant, and energy is transported to regions cooled by neutrinos, resulting in a total luminosity that can exceed the photon luminosity. 
To compare with the NS model, we also compute the neutrino luminosity from the accretion column using Eq. (1) of  \citet{Asthana:2023vvk}. At relatively low accretion rates  ($< 10^{20}\, \rm g\,s^{-1}$), the SS model predicts moderately higher neutrino fluxes for Swift J0243.6$+$6124, RX J0209.6$-$7427, and SMC X-3, exceeding the results of \citet{Asthana:2023vvk} by factors of $\sim 2-5$. In contrast, for ULXPs, the SS model yields neutrino fluxes that are comparable to or lower than those of the NS model, with typical reductions of $\lesssim 2$ and up to a factor of $\sim 5$ for extreme sources such as NGC 5907 ULX-1. Furthermore, the luminosity associated with heat transport along the SS surface is estimated to be $ L_{\rm heat} \sim 1.2 \times 10^{36}\, \rm erg\, s^{-1}$, independent of the mass accretion rate.

Within the framework of the SS model, our results indicate that the predicted neutrino flux at Earth is extremely low due to large distances for most extragalactic sources, with the notable exception of Swift J0243.6$+$6124, the closest ULXP, where the neutrino flux could reach $ F_{\nu} \sim 2.3 \times 10^{-2}\, \rm MeV \, cm^{-2}\, s^{-1}$. However,  even under optimistic parameter choices, the predicted fluxes remain below the diffuse MeV neutrino background. Related discussions of neutrino emission from accreting neutron stars can also be found in the literature, for example in \citet{Asthana:2023vvk}.

Concerning prospects for point-source detection, the quantitative results shown in Fig. \ref{fig: spectrum} indicate that the expected neutrino fluxes from Swift J0243.6$+$6124, RX J0209.6$-$7427, NGC 5907 ULX-1, and M51 ULX-8 typically lie one to two orders of magnitude below the diffuse background level in the relevant energy range. Therefore, even for the closest and brightest systems currently known, the signal would be strongly background-dominated. A rough implication of our results is that only sources at distances of order $\lesssim 1 \, \rm kpc$, and with comparably high luminosities, might approach the threshold of potential detectability, although such a detection remains highly challenging with present instruments.

This study highlights the importance of considering the detailed structure of the thermal mound and EOS of SSs in modeling high-luminosity accretion flows. While photon emission dominates at low accretion rates, neutrino emission can become the primary cooling channel at extreme accretion rates, potentially affecting the interpretation of observed ULXP luminosities.

\section*{Acknowledgements}

This work was supported by the China Postdoctoral Science Foundation (2024M760081), the National Natural Science Foundation of China (12447148, 12041301, 12121003 and 123B2045), and the National SKA Program of China (2020SKA0120100).


\section*{Data Availability}

The data underlying this paper will be shared on reasonable request to the
corresponding authors.



\bibliographystyle{mnras}
\bibliography{Neutrino_ULXPs} 

\bsp	
\label{lastpage}
\end{document}